\documentclass[11pt]{article}

\makeatletter 
\def\flE{\begin{picture}(0,0) 
   \put( 0.25,    0){\vector( 1, 0){0.50}} 
   \@ifstar{\@flE}{\@@flE}} 
\def\@flE  #1{\put( 0.5 ,-0.03){\makebox(0,0)[ t]{$#1$}}\end{picture}} 
\def\@@flE #1{\put( 0.5 , 0.03){\makebox(0,0)[ b]{$#1$}}\end{picture}} 
\def\flNE{\begin{picture}(0,0) 
   \put( 0.18, 0.18){\vector( 1, 1){0.64}} 
   \@ifstar{\@flNE}{\@@flNE}} 
\def\@flNE #1{\put( 0.52, 0.48){\makebox(0,0)[tl]{$#1$}}\end{picture}} 
\def\@@flNE#1{\put( 0.48, 0.52){\makebox(0,0)[br]{$#1$}}\end{picture}} 
\def\flN{\begin{picture}(0,0) 
   \put(    0, 0.20){\vector( 0, 1){0.60}} 
   \@ifstar{\@flN}{\@@flN}} 
\def\@flN  #1{\put( 0.03, 0.5 ){\makebox(0,0)[ l]{$#1$}}\end{picture}} 
\def\@@flN #1{\put(-0.03, 0.5 ){\makebox(0,0)[ r]{$#1$}}\end{picture}} 
\def\flNW{\begin{picture}(0,0) 
   \put(-0.18, 0.18){\vector(-1, 1){0.64}} 
   \@ifstar{\@flNW}{\@@flNW}} 
\def\@flNW #1{\put(-0.48, 0.52){\makebox(0,0)[bl]{$#1$}}\end{picture}} 
\def\@@flNW#1{\put(-0.52, 0.48){\makebox(0,0)[tr]{$#1$}}\end{picture}} 
\def\flW{\begin{picture}(0,0) 
   \put(-0.25,    0){\vector(-1, 0){0.50}} 
   \@ifstar{\@flW}{\@@flW}} 
\def\@flW  #1{\put(-0.5 , 0.03){\makebox(0,0)[ b]{$#1$}}\end{picture}} 
\def\@@flW #1{\put(-0.5 ,-0.03){\makebox(0,0)[ t]{$#1$}}\end{picture}} 
\def\flSW{\begin{picture}(0,0) 
   \put(-0.18,-0.18){\vector(-1,-1){0.64}} 
   \@ifstar{\@flSW}{\@@flSW}} 
\def\@flSW #1{\put(-0.52,-0.48){\makebox(0,0)[br]{$#1$}}\end{picture}} 
\def\@@flSW#1{\put(-0.48,-0.52){\makebox(0,0)[tl]{$#1$}}\end{picture}} 
\def\flS{\begin{picture}(0,0) 
   \put(    0,-0.2 ){\vector( 0,-1){0.60}} 
   \@ifstar{\@flS}{\@@flS}} 
\def\@flS  #1{\put(-0.03,-0.5 ){\makebox(0,0)[ r]{$#1$}}\end{picture}} 
\def\@@flS #1{\put( 0.03,-0.5 ){\makebox(0,0)[ l]{$#1$}}\end{picture}} 
\def\flSE{\begin{picture}(0,0) 
   \put( 0.18,-0.18){\vector( 1,-1){0.64}} 
   \@ifstar{\@flSE}{\@@flSE}} 
\def\@flSE #1{\put( 0.48,-0.52){\makebox(0,0)[tr]{$#1$}}\end{picture}} 
\def\@@flSE#1{\put( 0.52,-0.48){\makebox(0,0)[bl]{$#1$}}\end{picture}} 
\def\capsa(#1,#2)#3{\put(#1,#2){\makebox(0,0){$#3$}}} 
\def\diagr{\@ifnextchar [{\@diagr}{\@diagr[15ex]}} 
\def\@diagr[#1](#2,#3){\begingroup 
   \setlength{\unitlength}{#1} 
   \begin{picture}(#2,#3)} 
\def\enddiagr{\end{picture} 
   \endgroup} 
\def\indiag{\@ifnextchar [{\@indiag}{\@indiag[15ex]}} 
\def\@indiag[#1](#2,#3){\begingroup 
   \setlength{\unitlength}{#1} 
   \medskip 
   \begin{center} 
   \begin{picture}(#2,#3)} 
\def\exdiag{\end{picture} 
   \end{center} 
   \medskip 
   \endgroup} 
\makeatother

\makeatletter
\def\fldins{\begin{picture}(0,0)
         \put(0.25,0.03){\vector(3,1){1}}   
         \@ifstar{\@fldins}{\@@fldins}}
\def\@fldins#1{\put(0.75,0.18){\makebox(0,0)[tl]{$#1$}}\end{picture}}  
\def\@@fldins#1{\put(0.75,0.22){\makebox(0,0)[br]{$#1$}}\end{picture}} 
\def\flddins{\begin{picture}(0,0)
         \multiput(0.25,0.03)(0.286,0.095){3}{\line(3,1){0.143}}  
         \put(1.113,0.315){\vector(3,1){0.143}}   
         \@ifstar{\@fldins}{\@@fldins}}
\makeatother

\makeatletter

\def\qed{\ifvmode\removelastskip\fi
{\unskip\nobreak\hfil\penalty50\hbox{}\nobreak\hfil
\hbox{\vrule height1.2ex width1.2ex}\parfillskip=0pt
\finalhyphendemerits=0 \par\smallskip}}

\def\dif{{\rm d}}
\def\deriv{\@ifnextchar[{\@deriv}{\@deriv[]}}
   \def\@deriv[#1]#2#3{\mathchoice%
{{\dif^{#1}#2\over\dif{#3}^{#1}}}{{\dif^{#1}#2/\dif{#3}^{#1}}}%
{{\dif^{#1}#2\over\dif{#3}^{#1}}}{{\dif^{#1}#2/\dif{#3}^{#1}}}}
\def\derpar#1#2{\mathchoice%
{{\partial#1\over\partial#2}}{{\partial#1/\partial#2}}%
{{\partial#1\over\partial#2}}{{\partial#1/\partial#2}}}

\def\restric#1#2{{\left. #1 \right|_{#2}}}
\def\secteqno{\@addtoreset{equation}{section}%
\def\theequation{\thesection.\arabic{equation}}}
\newcounter{subequation}
\def\thesubequation{\alph{subequation}}
\def\sneqnarray{\stepcounter{equation}\let\@currentlabel=\theequation
\setcounter{subequation}{1}
\def\@eqnnum{{\rm (\theequation.\thesubequation)}}
\global\@eqcnt\z@\tabskip\@centering\let\\=\@eqncr\let\@@eqncr=\@@sneqncr
$$\halign to \displaywidth\bgroup\@eqnsel\hskip\@centering
 $\displaystyle\tabskip\z@{##}$&\global\@eqcnt\@ne
 \hskip 2\arraycolsep \hfil${##}$\hfil
 &\global\@eqcnt\tw@ \hskip 2\arraycolsep $\displaystyle\tabskip\z@{##}$\hfil
  \tabskip\@centering&\llap{##}\tabskip\z@\cr}
\def\endsneqnarray{\@@sneqncr\egroup $$\global\@ignoretrue}
\def\@@sneqncr{\let\@tempa\relax
   \ifcase\@eqcnt \def\@tempa{& & &}\or \def\@tempa{& &}
   \else \def\@tempa{&}\fi
     \@tempa \if@eqnsw\@eqnnum\stepcounter{subequation}\fi
     \global\@eqnswtrue\global\@eqcnt\z@\cr}
\def\nobiblabels{\def\@lbibitem[##1]##2{\@bibitem{##2}}}
\makeatother
\def\tabaddress#1{{\small\it\begin{tabular}[t]{c}#1\\[1.2ex]\end{tabular}}}
\def\artit#1{``#1'',}
\def\ben{\begin{enumerate}}
\def\een{\end{enumerate}}
\def\beq{\begin{equation}}
\def\eeq{\end{equation}}
\def\bea{\begin{eqnarray}}
\def\eea{\end{eqnarray}}
\def\beann{\begin{eqnarray*}}
\def\eeann{\end{eqnarray*}}
\def\beasn{\begin{sneqnarray}}
\def\eeasn{\end{sneqnarray}}

\newtheorem{teor}{Theorem}
\newtheorem{prop}{Proposition}

\newenvironment{defin}{\paragraph{Definition}}{\removelastskip\smallskip}

\let\ds=\displaystyle
\def\buildord#1\over#2{\mathord{\mathop{\kern0pt #2}\limits^{#1}}}

\def\Id{{\rm Id}}

\def\Real{{\bf R}}
\let\isom=\cong

\def\Ker{\mathop{\rm Ker}\nolimits}
\def\transp#1{{}^{t}\kern-.15em\relax#1}
\def\Img{\mathop{\rm Im}\nolimits}

\def\feble#1{\mathrel{\mathop{\simeq}\limits_{#1}}}
\def\scomp{\bullet}

\def\lop{\!\cdot\!}

\def\Tan{{\rm T}}

\def\Lio{\Delta}

\def\fin{{\rm f}}

\def\proof{\noindent{\it Proof}.\quad}

\let\eps=\varepsilon
\def\w{{\sf\bf w}}
\def\comp{\mathbin{\scriptstyle\circ}}
\def\scomp{\mathbin{\scriptstyle\bullet}}
\def\Fl{{\rm F}}
\def\Lie{{\cal L}}

\def\UPCMAT{Departament de Matem\`atica Aplicada IV\\
   Universitat Polit\`ecnica de Catalunya\\
   Campus Nord UPC, edifici C3\\
   C.~Jordi Girona, 1\\
   08034 Barcelona\\
   Catalonia, Spain}

\def\UBECM{Departament d'Estructura i Constituents de la Mat\`eria\\
   Universitat de Barcelona\\
   Av.~Diagonal, 647\\
   08028 Barcelona\\
   Catalonia, Spain}

\title{\sf
Symmetries and infinitesimal symmetries\\
of singular differential equations}

\author{\sf
Xavier Gr\`acia$^a$ and Josep M. Pons$^b$
\\[2mm]
\tabaddress{$^a$\UPCMAT}
\\
\tabaddress{$^b$\UBECM}
\\[2mm]
\small\sf e-mails: 
xgracia@mat.upc.es, 
pons@ecm.ub.es
}

\date{\sf 15 January 2002}

\secteqno
\begin{document}
\abovedisplayskip=6pt plus 3pt minus 1pt
\belowdisplayskip=\abovedisplayskip
\belowdisplayshortskip=4pt plus 3pt minus 1pt

\maketitle
\thispagestyle{empty}

\begin{abstract}
\parindent 0pt
\noindent

The aim of this paper is to study symmetries 
of linearly singular differential equations,
namely,
equations that can not be written in normal form
because the derivatives are multiplied by a singular linear operator. 

The concept of geometric symmetry 
of a linearly singular differential equation is introduced
as a transformation that preserves the geometric data that define the problem. 
It is proved that such symmetries 
are essentially equivalent to dynamic symmetries, 
that is, 
transformations mapping solutions into solutions. 
Similar results are given for infinitesimal symmetries. 

To study the invariance of several objects under the flows of vector fields, 
a careful study of infinitesimal variations is performed, 
with a special emphasis on infinitesimal vector bundle automorphisms.



PACS 1999: 02.30.Hq, 02.40.Vh

\end{abstract}

\newpage
\section{Introduction}

\def\Path{\hbox{$\cal P$}}
\def\Sol{\hbox{$\cal S$}}

Let $M$ be a manifold, and $\Path$ the set of paths in~$M$.
Consider a subset $\Sol \subset \Path$, which may be understood as the set of
{\it solutions} of a certain {\it problem} on the set of paths.
This problem ---usually we think of a differential equation--- 
may be stated in terms of several objects, 
the {\it data} of the problem. 
The solutions  of the problem are the paths in $\Path$ 
that satisfy some condition with respect to the data 
that identify the problem. 
Examples of data can be 
a vector field, a submanifold, a connection, a potential function, a
lagrangian, etc, 
and for each case 
a specification of the associated problem must be given: 
let it be the search for the integral curves of a vector field, 
the critical paths of an action functional, etc.

Since the {\it problem} is identified by the {\it data} and solved by the  
{\it solutions}, 
there appear naturally two different concepts of what a symmetry of 
the problem is, 
differing as to whether the emphasis is put on the side of the {\it data} 
or on the side of the {\it solutions}. 
In a certain sense, a transformation that preserves the data 
is a symmetry of the problem; 
and so it is, but in another sense,  
a transformation that maps solutions into solutions. 
In order to avoid confusion,
we can call {\it geometric} the former symmetries, 
and {\it dynamic} the latter. 
In fact, the word {\it dynamic} has the reminiscence of 
the equations of motion that set the dynamics of a physical problem. 
Also, in most cases, the data of the problem will have {\it geometric}
significance, and hence the name suggested. 
Usually a geometric symmetry of a problem 
will also be a dynamic symmetry;
thus the search for geometric symmetries  
will be a relevant part of a wider subject: 
the search for dynamic symmetries.  
Noether transformations of an action functional are an example of this. 

As for the type of transformations of the paths,
we will confine ourselves to 
{\it point transformations} (of the dependent variable),
which arise as $\gamma \mapsto \varphi_*[\gamma] := \varphi \comp \gamma$,
for a certain diffeomorphism $\varphi \colon M \to M$.
(Among non-point transformations we have for instance
reparametrisations of the independent variable,
and the generalised transformations,
where the transformation involves also the derivatives of the path
---see
\cite{Olv-diffeq}.)

\medskip

In this paper we will deal with problems resulting in first-order autonomous 
differential equations on~$M$. 
Among the possible data that identify the problem, 
there is the differential equation itself, 
considered as an implicit relation involving 
a path $x(t)$ and its derivative with respect to the evolution parametre:
$$
F(x,\dot x) = 0 .
$$
Equations of this form are often called 
differential-algebraic, or implicit differential equations.

Of course, 
if one can isolate the derivative,
$$
\dot x = f(x) ,
$$
the equation is said to be in normal form,
and giving an initial condition 
$(t_0,x(t_0))$
determines uniquely the solution $x(t)$.
However, we are mainly interested in the singular case.

More precisely, 
we are interested in implicit differential equations of the form
$$
A(x) \dot x = b(x) ,
$$
where the velocities can not be isolated because of 
an everywhere singular linear operator $A(x)$ multiplying them.
Such equations may be called linearly singular differential equations.
This general class of implicit differential equations 
was geometrically presented in 
\cite{GP-unif}
\cite{GP-gener}.
In these papers it is pointed out that
many interesting systems of mathematical physics and
applied mathematics are linearly singular.

In more detail, this framework describes the equations of motion of  
the presymplectic dynamical systems
\cite{GNH-pres}
(including their applications to lagrangian and hamiltonian mechanics
\cite{Dir-lectures}
\cite{GN-pres-lag}
\cite{MT-ham}
\cite{Ski-mixt}
\cite{SR-mixt}),
the first-order lagrangian formalism
\cite{GP-gener},
the higher-order lagrangian dynamics
\cite{GPR-higher}
\cite{LR-higher}
(including also their ``higher order differential equation'' conditions
\cite{GPR-higcond}),
and systems with nonholonomic constraints.

In addition to these applications of interest for mathematical physics,
one can find applications of 
implicit and linearly singular differential equations
to electrical and chemical engineering, control theory, economics, etc 
(see for more details examples and references in 
\cite{GP-gener} 
\cite{GMR-nantes95} 
\cite{HLR-89} 
\cite{Rhe-84}). 
As well as all the mentioned papers,
there are also many articles and books studying 
geometric features 
\cite{CO-88}
\cite{HB-84}
\cite{MMT-92} 
\cite{MMT-95} 
\cite{MR-implicit}
\cite{MT-ham}
\cite{Rei-90}
\cite{Rei-91}
\cite{RR-94}
\cite{Tak-76}
and numerical methods 
\cite{Cam-singeq} 
\cite{HLR-89}
\cite{HW-91} 
for implicit equations.

The symmetries of an implicit differential equation
can be studied using general techniques 
\cite{Olv-diffeq}.
Important topics as lagrangian systems and presymplectic systems 
have been widely studied 
(see for instance
\cite{Olv-diffeq} 
\cite{LM-96} 
and references therein).
Besides these cases, 
there are few references on 
symmetries of implicit differential equations:
we could point out the articles
\cite{MMT-92}, 
where symmetries and constants of motion for implicit systems 
$F(x,\dot x)=0$ are studied;
\cite{CO-88}, 
which contains a study of normal forms of linearly singular systems 
given by a vector bundle morphism 
$A \colon \Tan M \to \Tan M$; 
\cite{MR-implicit},
which deals with symmetries of linearly singular systems 
given by a vector bundle morphism 
$A \colon \Tan M \to \Tan^*M$;
and the recent study
\cite{BS-01} 
of symmetries and reduction of Dirac structures.

The main purpose of this article is to study the symmetries of 
a linearly singular differential equation. 
For such an equation we can consider 
the geometric symmetries preserving the data ($A$, $b$) 
that define the equation.
It will be proved that
any dynamic symmetry of the differential equation
may be locally realised as a geometric symmetry of the data.
A similar result will be also given for infinitesimal symmetries.

\medskip

The paper is organised as follows.
Section~2 presents the geometric framework of
linearly singular differential equations, and gives several basic results.
Section~3 studies the symmetries of such a system,
and relates them to the symmetries of the
associated implicit differential equation.
In section~4 the concept of infinitesimal symmetry is presented,
and a study similar to that of section~3 is performed.
Section~5 particularises all the results to regular and to consistent systems. 
Section~6 is devoted to an example, and section~7 to conclusions.
Finally, there is an appendix dealing with calculus of
infinitesimal variations,
and more particularly to the invariance of maps 
under the action of flows of vector fields,
and to infinitesimal symmetries of vector bundles.

The tools used in this paper are those of differential geometry, 
in particular 
manifolds and submanifolds, vector fields and their flows, 
and vector bundles and their morphisms
\cite{AMR-manif} 
\cite{Die-ea3}
\cite{KMS-natural}.
Throughout the paper the manifolds are finite-dimensional and paracompact,
and the maps are smooth;
``differential equation'' means
``first-order autonomous ordinary differential equation''.

\section{Linearly singular differential equations}
\label{sec-gener}

In this section we recall some of the main results from
\cite{GP-unif}
\cite{GP-gener},
and we give additional results to be used later on.

\subsection{The geometric framework}

\begin{defin}
An {\it implicit system}\/ 
on a manifold~$M$ is a submanifold 
$D \subset \Tan M$. 
It defines an {\it implicit differential equation}, 
for which a path $\xi \colon I \to M$ is a {\it solution}\/ when 
its lift to the tangent bundle, $\dot \xi$, is contained in~$D$:
\beq
\dot\xi(I) \subset D .
\label{eqmov-D}
\eeq
\end{defin}

When $D$ is the image of a vector field $X$ on~$M$, $D=X(M)$,
one has an {\it explicit}\/ differential equation,
or says that the equation can be put in normal form.
Then a path $\xi$ is a solution of~$D$ iff $\dot\xi = X \comp \xi$.

\begin{defin}
(\cite{GP-unif}
\cite{GP-gener})
A {\it linearly singular system} 
is a quintuple $(M,F,\pi,A,b)$ given by
a manifold~$M$, 
a vector bundle $\pi \colon F \to M$,
a vector bundle morphism $A \colon \Tan M \to F$,
and a section $b \colon M \to F$. 
These data define a {\it linearly singular differential equation}, 
for which a {\it solution}\/ is a path $\xi \colon I \to M$ such that
\beq
A \comp \dot\xi = b \comp \xi .
\label{eqmov-xi}
\eeq
The {\it associated implicit system}\/ is the subset
\beq
D = A^{-1} (b(M)) \subset  \Tan M .
\eeq
\end{defin}

We will use the notation $(A \colon \Tan M \to F,b)$
to refer to a linearly singular system.
The following diagram shows all these data:
$$
\diagr(2,1.15)
\capsa(0,0){I}
\capsa(1,0){M}
\capsa(1,1){\Tan M}
\capsa(0,0){\flE{\xi}}
\capsa(0,0){\flNE{\dot \xi}}
\capsa(1,1){\flS{\tau_M}}
\capsa(2,1){F}
\capsa(2,1){\flSW*{\pi}}
\capsa(1.05,0){\flNE*{b}}
\capsa(1,1){\flE{A}}
\enddiagr
$$

\begin{prop} 
The differential equations defined by 
a linearly singular system 
\linebreak
$(A \colon \Tan M \to F,b)$ 
and its associated implicit system~$D$ 
have the same solutions.
\end{prop}
\proof
Equation (\ref{eqmov-D}) means that, for each~$t$,
$\dot\xi(t) \in D$. 
This is equivalent to 
$A \lop \dot\xi(t) \in b(M)$,
and being $A$ fibre-preserving this is equivalent to 
$A \lop \dot\xi(t) = b(\xi(t))$.
\qed

Note that in general (\ref{eqmov-xi}) may not have solutions passing
through every point in~$M$,
and if there is a solution passing through a point $x$ at a given time
it may not be unique.
We call {\it motion set}\/ the set $S \subset M$ of points
by which a solution passes.

\medskip

It is useful to try 
to describe the solutions of the equation of motion (\ref{eqmov-xi})
as integral curves of vector fields.
More precisely,
if $M' \subset M$ is a submanifold
and $X$ is a vector field on $M$ {\it tangent}\/ to~$M'$,
then the integral curves of $X$ contained in $M'$ are solutions
of the equation of motion (\ref{eqmov-xi})
if and only if $X$ satisfies
\beq
A \comp X \feble{M'} b ,
\label{eqmov-X}
\eeq
where the notation $\feble{M'}$ means equality at the points of~$M'$.
Let us remark that this is an equation both for
$X$ and $M'$,
since in general there will not be a vector field 
satisfying this equation all over~$M$.

\subsection{The constraint algorithm}

Consider a linearly singular system $(A \colon \Tan M \to F,b)$.
To solve the corresponding differential equation 
a consistency algorithm may be performed.
This algorithm is indeed a generalisation of 
the presymplectic constraint algorithm
\cite{GNH-pres},
which is a geometrisation of Dirac's theory for singular lagrangians
\cite{Dir-lectures}.
Let us describe this algorithm briefly.

\begin{defin}
The {\it primary constraint subset}\/ is the set $M_1 \subset M$
of points $x$ where the linear equation
$A_x \lop u_x = b(x)$
is consistent:
\beq
M_1 =
\{ x \in M  \mid  b(x) \in \Img A_x \} ,
\label{M1}
\eeq
The functions of $M$ vanishing on $M_1$
constitute the ideal of {\it primary constraints}.
\end{defin}

The reason for the terminology is clear:
in view of the differential equation of motion (\ref{eqmov-xi}),
if $\xi$ is a (smooth) solution of it,
then necessarily $\xi$ lives in~$M_1$.
As for the constraints,
in this paper we are not especially interested in 
explicit procedures to compute them; see
\cite{GP-gener}
for more details.

To proceed further it is convenient to require some regularity conditions
on $A$ and~$M_1$:

\begin{defin}
We will refer to the {\it regularity assumption}\/ as the following conditions
to be satisfied by a linearly singular system $(A \colon \Tan M \to F,b)$:
\ben
\itemsep 0pt plus 1pt
\item
The morphism $A$ has constant rank
---thus $\Ker A \subset \Tan M$ and $\Img A \subset F$ 
are vector subbundles,
$D \subset \Tan M$ is a closed submanifold 
and $M_1$ is a closed subset.
\item
The primary constraint subset $M_1 \subset M$ is a non-empty submanifold.
\een
\end{defin}

Let us assume that our linearly singular system 
satisfies the regularity assumption, 
and let $\xi$ be a solution of the corresponding differential equation. 
It is clear that 
$\xi$ is a also a solution of the linearly singular differential equation
defined by restricting all the problem to~$M_1$,
namely, the {\it subsystem}
$(A_1 \colon \Tan M_1 \to F_1,b_1)$,
where $F_1 = \restric{F}{M_1}$,
and $A_1$ and $b_1$
are the corresponding restrictions to~$M_1$.

Note that the problem is not yet solved: 
for a point $x \in M_1$,
$b(x)$ does not necessarily belong to the image of
the restriction of $A_x$ to the subspace $\Tan_x M_1 \subset \Tan_x M$. 
Repeating the consistency analysis for the subsystem
yields a subset $M_2 := (M_1)_1 \subset M_1$.

Let us {\it assume}\/ that 
the regularity assumption holds for the successive subsystems. 
The repetition of the consistency analysis on the subsystems 
yields an algorithm that reaches  
---in a finite number of steps since $M$ is finite-dimensional---  
a {\it final constraint submanifold} 
$\ds
M_\fin := \bigcap_{i \geq 0} M_i .
$
The solutions of the original problem
are the solutions of the equation of motion of the linearly singular system
$(A_\fin \colon \Tan M_\fin \to F_\fin,b_\fin)$
defined by restriction to $M_\fin$.
By construction,
$b_\fin$ has its image in $\Img A_\fin$
---otherwise the algorithm would not be finished.
Therefore the equation
\beq
A_\fin \comp X_\fin = b_\fin
\label{eqmov-final}
\eeq
for a vector field $X_\fin$ in $M_\fin$ has solutions.
Since $M_\fin$ is closed
these solutions can be extended throughout $M$
to yield solutions $X$ of the equation of motion
(\ref{eqmov-X}) along $M' = M_\fin$
which are tangent to this submanifold.
Given a particular solution $X_\fin$ of (\ref{eqmov-final}),
the set of solutions is $X_\fin + \Ker A_\fin$.
Therefore there is a unique solution (on~$M_\fin$)
iff $A_\fin$ is injective.
$$
\diagr(2,1.15)
\capsa(0,0){M_\fin}
\capsa(-0.05,0){\flN{X_\fin}}
\capsa(0,1){\Tan M_\fin}
\capsa(0,1){\flS{}}
\capsa(1,1){\Img A_\fin}
\capsa(0,0){\flNE*{b_\fin}}
\capsa(0,1){\flE{A_\fin}}
\capsa(2,1){F_\fin}
\capsa(1,1){\flE{}}
\enddiagr
$$
So note that the final dynamics is simply 
that of a linearly singular system 
where the morphism $A$ is surjective.

Note finally that if the $M_i$ fail to be submanifolds,
then in order to apply the constraint algorithm
some points of the base space $M$ may have to be removed;
in this case only a subset $M_\fin \subset S$ of the motion set 
will be obtained,
and the motion set $S$ may not be a submanifold
---see some examples in
\cite{GP-gener}.

\subsection{Morphisms of linearly singular systems}

\begin{defin}
A {\it morphism of linearly singular systems}\/
between $(A \colon \Tan M \to F,b)$ and $(A' \colon \Tan M' \to F',b')$
is a morphism $(\varphi,\Phi)$
between the vector bundles $F \to M$ and $F' \to M'$ 
(so it satisfies
$
\varphi \comp \pi = \pi' \comp \Phi 
$)
such that
\beasn
\Phi \comp b      &=&  b' \comp \varphi ,
\\
\Phi \comp A      &=&  A' \comp \Tan \varphi .
\label{mor}
\eeasn
\end{defin}

Let us show all this in a diagram:
$$
\diagr(2.5,1.55)
\capsa(0,0){M}
\capsa(0,1){\Tan M}
\capsa(0,1){\flS*{}}
\capsa(1,1){F}
\capsa(1,1){\flSW*{\pi}}
\capsa(0.05,0){\flNE*{b}}
\capsa(0,1){\flE*{A}}

\capsa(1.5,0.4){M'}
\capsa(1.5,1.4){\Tan M'}
\capsa(1.5,1.4){\flS*{}}
\capsa(2.5,1.4){F'}
\capsa(2.5,1.4){\flSW*{\pi'}}
\capsa(1.55,0.4){\flNE*{b'}}
\capsa(1.5,1.4){\flE{A'}}

\capsa(0,0){\fldins*{\varphi}}
\capsa(0,1){\fldins{\Tan \varphi}}
\capsa(1,1){\fldins*{\Phi}}
\enddiagr
$$

With this definition the linearly singular systems 
constitute a category. 
Its isomorphisms correspond to the case when 
$(\varphi,\Phi)$ is an isomorphism of vector bundles. 
In this case, in general we can define 
\beq
\Phi_*[A] := \Phi \comp A \comp (\Tan \varphi)^{-1} , 
\qquad 
\Phi_*[b] := \Phi \comp b \comp \varphi^{-1} ,
\label{push}
\eeq
where $\Phi_*$ denotes the push-forward through the isomorphism
$(\varphi,\Phi)$; 
then the condition to be an isomorphism of linearly singular systems is 
$\Phi_*[A] = A'$, $\Phi_*[b] = b'$. 

Note also the following trivial remark: 
if $\Phi \colon F \to F$ is a base-preserving automorphism, 
it defines an isomorphism between 
$(A \colon \Tan M \to F,b)$ and 
$(\Phi \comp A \colon \Tan M \to F,\Phi \comp b)$. 
This reflects the fact that the equations 
$A \comp \dot\xi = b \comp \xi$ and 
$(\Phi \comp A) \comp \dot\xi = (\Phi \comp b) \comp \xi$ 
are completely equivalent. 

It is easily proved that a morphism applies solutions of the corresponding 
differential equation into solutions. 
Other constructions with linearly singular systems can be carried out: 
subsystems, quotients, products, \dots\ 
These constructions induce natural morphisms. 
See
\cite{GP-gener}
for more details.

\subsection{Primary dynamical vector fields}

Let us have a closer look to the first stage of the constraint algorithm.

\begin{prop}
Consider a linearly singular system $(A \colon \Tan M \to F,b)$.
Then:
\ben
\itemsep 0pt plus 1pt
\item
$M_1 = \tau_M(D)$,
where $\tau_M \colon \Tan M \to M$ is the natural projection.
\item
If the regularity assumption is satisfied,
$D \subset \restric{\Tan M}{M_1} \to M_1$
is an affine subbundle
modelled on $\restric{(\Ker A)}{M_1}$.
\een
\label{prop-Dafi}
\end{prop}
\proof 
For the first assertion,
$x \in M_1$ iff 
there exists $v_x \in \Tan_xM$ such that $A \lop v_x = b(x)$,
which is equivalent to saying that
$v_x \in D_x$
---we write as usual $D_x = D \cap \Tan_xM$.

Now consider the restriction of~$A$ to the submanifold~$M_1$,
$A_1 \colon \restric{\Tan M}{M_1} \to \restric{F}{M_1}$.
Since the section $b_1 = \restric{b}{M_1}$ is in the image of~$A_1$
and $A_1$ is a vector bundle morphism with constant rank,
we have that
$D = A_1^{-1}(b_1(M_1))$ is an affine subbundle of $\restric{\Tan M}{M_1}$
modelled on $\Ker A_1 = \restric{(\Ker A)}{M_1}$.
\qed

\begin{defin}
A section of~$D \to M_1$ is called a 
{\it primary dynamical vector field}.
\end{defin}

Of course we can suppose that such a section 
is extended to a vector field $X$ on~$M$.
Then, saying that $X$ is a primary vector field means that
\beq
A \comp X \feble{M_1} b .
\label{camp-primari}
\eeq
Such vector fields constitute a first approach to the final dynamics
(the tangency of $X$ to $M_1$ is not guaranteed).

If $X_o$ is a primary dynamical vector field,
another vector field $X$ is a primary field if and only if
it differs from $X_o$ on a section of $\Ker A$ on~$M_1$.
Thus, if $(\Gamma_\mu)_{1\le \mu\le m}$
is a local frame for $\Ker A$ near~$M_1$,
then there locally exist functions $g^\mu$,
uniquely determined on~$M_1$,
such that, locally,
\beq
X \feble M_1
X_o + \sum_\mu g^\mu \Gamma_\mu ;
\label{prim-prim}
\eeq
see 
\cite{GP-gener}
for more details on how an explicit computation of the final dynamics
can be obtained in this way.

\section{Symmetries of linearly singular systems}

We have pointed out in the introduction that 
one may define several concepts of symmetry
of a differential equation
according to the data that define it.
Our purpose now is to study the natural symmetries
of linearly singular systems. 

\begin{defin}
A {\it symmetry of an implicit system}\/
$D \subset \Tan M$ is
a diffeomorphism $\varphi \colon M \to M$ 
leaving $D$ invariant, that is,
\beq
(\Tan\varphi)(D) \subset D .
\eeq
\end{defin}

\begin{prop}
A symmetry $\varphi$ of $D$ maps solutions 
of the corresponding differential equation 
into solutions.
\end{prop}
\proof
It is immediate:
if $\xi$ is a solution 
(that is, $\dot\xi(t) \in D$)
then $\varphi\comp\xi$ is also, since
$(\varphi\comp\xi)^{\textstyle.}(t) 
=   (\Tan\varphi) \lop \dot\xi(t)
\in D$.
\qed

Though this is the natural geometric definition of a symmetry of~$D$,
this is not a necessary condition for $\varphi$ 
to define a symmetry of the solutions of the differential equation,
because an implicit differential equation may not have solutions
passing through every point in~$D$.
But with a convenient refinement of it, this condition
essentially characterises the symmetries of 
the solutions of the differential equation.
More precisely, following the terminology of
\cite{Olv-diffeq},
let us call $D$ {\it locally solvable}\/
if for each $v \in D$ 
there is a solution $\xi$ of the implicit differential equation 
such that $\dot \xi(0) = v$ 
(see also
\cite{MMT-92}).

\begin{prop}
Suppose that $D$ is locally solvable.
Then a diffeomorphism $\varphi \colon M \to M$ 
is a (geometric) symmetry of $D$ iff 
it is a (dynamic) symmetry of the corresponding differential equation. 
\end{prop} 
\proof
The direct implication is the preceding proposition. 
For the converse, let $v \in D$. 
For a certain solution~$\xi$, 
$\dot\xi(0)=v$, 
and since $\varphi\comp\xi$ is also a solution, 
$(\Tan\varphi) \lop v = (\varphi\comp\xi)^{\textstyle.}(0) \in D$.
\qed

When $D$ is not locally solvable, 
if one can perform a ``constraint algorithm''  
to pass to a locally solvable $D' \subset D$, 
then the dynamic symmetries of~$D$ (or~$D'$) 
are in correspondence with the geometric symmetries of~$D'$. 

\begin{defin}
A {\it symmetry of a linearly singular system}\/ 
$(A \colon \Tan M \to F,b)$ 
is an isomorphism with itself, 
that is, 
a vector bundle automorphism 
$(\varphi,\Phi)$ of $\pi \colon F \to M$ 
such that 
$$ 
b = \Phi_*[b], \quad  A = \Phi_*[A] . 
$$ 
\end{defin} 

We have already said that 
such a transformation maps solutions into solutions.
A more precise result is:

\begin{prop} 
Let $\varphi$ be the base map of a symmetry $(\varphi,\Phi)$ of a 
linearly singular system. 
Then $\varphi$ is a symmetry of the associated 
implicit system~$D$. 
\end{prop} 
\proof
We have to show that $D$ is $\varphi$-invariant. 
Let $v_x \in D$:
$A_x \lop v_x = b(x)$.
Then, according to (\ref{mor}),
$$
A_{\varphi(x)} \lop (\Tan_x(\varphi) \lop v_x) =
\Phi_x \lop (A_x \lop v_x) =
\Phi_x \lop b(x) =
b(\varphi(x)) ,
$$
which shows that $\Tan_x(\varphi) \lop v_x \in D$.
\qed

We want to prove a kind of converse of this statement.
To this end, first we state an auxiliary result:

\begin{prop}
Let $A \colon E \to F$ be a vector $B$-bundle morphism, and 
$A' \colon E' \to F'$ a vector $B'$-bundle morphism.
Let $S \colon E \to E'$ be a vector bundle isomorphism over a map
$\varphi \colon B \to B'$,
and such that
$S \lop \Ker A \subset \Ker A'$.
Suppose that 
$A$ and $A'$ have the same constant rank,
and that $F$ and $F'$ have the same rank.

Then {\em locally}\/
there exists a vector bundle isomorphism $T \colon F \to F'$
such that $A' \comp S = T \comp A$.
\label{prop-aux}

\end{prop}

So this proposition deals with the commutativity 
of the ``upper'' square of the following diagram
by means of a certain morphism~$T$:
$$
\diagr(2.5,1.55)
\capsa(0,0){B}
\capsa(0,1){E}
\capsa(0,1){\flS*{}}
\capsa(1,1){F}
\capsa(1,1){\flSW*{}}
\capsa(0,1){\flE*{A}}

\capsa(1.5,0.4){B'}
\capsa(1.5,1.4){E'}
\capsa(1.5,1.4){\flS*{}}
\capsa(2.5,1.4){F'}
\capsa(2.5,1.4){\flSW*{}}
\capsa(1.5,1.4){\flE{A'}}

\capsa(0,0){\fldins*{\varphi}}
\capsa(0,1){\fldins{S}}
\capsa(1,1){\flddins*{T}}
\enddiagr
$$
\proof
Since $S \colon E \to E'$ is an isomorphism,
the hypotheses on the kernels and on the ranks imply that
$S (\Ker A) = \Ker A'$, 
and therefore $S$ defines an isomorphism
$\bar S \colon E/ \Ker A \to E' / \Ker A'$. 
Using the canonical isomorphisms 
$E/ \Ker A \cong \Img A$, 
$\bar S$ defines an isomorphism
$T_0 \colon \Img A \to \Img A'$.

Since any vector subbundle is a direct factor,
we can put $F = \Img A \oplus F_1$ and $F' = \Img A' \oplus F_1'$.
Any vector bundle map $T_1 \colon F_1 \to F_1'$
can be combined with $T_0$ to obtain a vector bundle morphism~$T$
such that $T \comp A = A' \comp S$.
The condition on the ranks {\it locally}\/ allows to choose $T_1$
to be an isomorphism,
and therefore~$T$.
\qed

Note however that $F_1$ and $F_1'$ in the proof need not be isomorphic,
even if $F$ and~$F'$ are so.
Therefore the last assertion in the proposition is necessarily local.
The proof also shows that $T$ is uniquely defined only on $\Img A$.

Now we are ready for the main result:

\begin{teor}
Let $(A \colon \Tan M \to F,b)$ be a linearly singular system 
satisfying the regularity assumption, 
and let $D = A^{-1}(b(M))$ be the associated implicit system. 

Let $\varphi \colon M \to M$ be a diffeomorphism. 
The following statements are equivalent: 
\ben
\itemsep 0pt plus 1pt
\item
$\varphi$ is a symmetry of the implicit system  
$D \subset \Tan M$.
\item
The restriction of 
$\Tan \varphi (\Ker A)$ to $M_1$ is in $\Ker A$ and, 
for any primary vector field~$X$, 
the restriction of $\varphi_*[X] - X$ to $M_1$ is in $\Ker A$. 
\item
$\varphi$ is {\em locally} the base map of a symmetry of the 
linearly singular system. 
\een
\label{teor-main}
\end{teor}
\proof
Condition (1) means that
\beq
\Tan_x(\varphi) \lop D_x = D_{\varphi(x)}
\label{sym1}
\eeq
for each $x \in M_1$.
Thanks to the affine structure of $D \to M_1$
(proposition~\ref{prop-Dafi}),
$$
\restric{D}{M_1} = \restric{X}{M_1} + \restric{\Ker A}{M_1} ,
$$
where $X$ is a primary dynamical vector field on~$M$.
At $x$ we have
$$
D_x = X(x) + \Ker A_x .
$$
So by condition (1) we have
\beasn
&& \Tan_x(\varphi) \cdot \Ker A_x = \Ker A_{\varphi(x)} 
\\
&& \Tan_x(\varphi) \cdot X(x) - X(\varphi(x)) \in \Ker A_{\varphi(x)} 
\label{sym2}
\eeasn
thus obtaining condition~(2),
and conversely.

Now let us apply proposition~\ref{prop-aux} to the following diagram:
$$
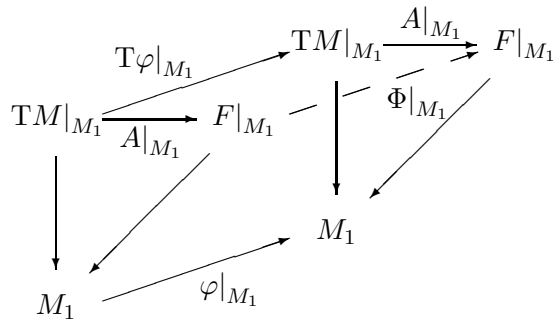
\diagr(2.5,1.55)
\capsa(0,0){M_1}
\capsa(0,1){\restric{\Tan M}{M_1}}
\capsa(0,1){\flS*{}}
\capsa(1,1){\restric{F}{M_1}}
\capsa(1,1){\flSW*{}}
\capsa(0,1){\flE*{\restric{A}{M_1}}}

\capsa(1.5,0.4){M_1}
\capsa(1.5,1.4){\restric{\Tan M}{M_1}}
\capsa(1.5,1.4){\flS*{}}
\capsa(2.5,1.4){\restric{F}{M_1}}
\capsa(2.5,1.4){\flSW*{}}
\capsa(1.5,1.4){\flE{\restric{A}{M_1}}}
\capsa(0,0){\fldins*{\restric{\varphi}{M_1}}}
\capsa(0,1){\fldins{\restric{\Tan \varphi}{M_1}}}
\capsa(1,1){\flddins*{\restric{\Phi}{M_1}}}
\enddiagr
$$
%
We conclude the local existence of a vector bundle isomorphism
$\restric{\Phi}{M_1}$ over $\restric{\varphi}{M_1} \colon M_1 \to M_1$
closing the ``upper square'' in the diagram.
We extend $\restric{\Phi}{M_1}$ locally to a vector bundle morphism
$\Phi \colon F \to F$,
which we can assure to be an isomorphism at least on an open 
neighbourhood of~$M_1 \subset M$.
We have
\beasn
&& \Phi_x \comp A_x = A_{\varphi(x)} \comp \Tan_x(\varphi) ,
\label{sym3}
\eeasn
which means that $\Phi_*[A] = A$.
Moreover,
$$
\Phi_x \cdot b(x) =
\Phi_x \cdot A_x \cdot X(x) =
A_{\varphi(x)} \cdot \Tan_x(\varphi) \cdot X(x) =
A_{\varphi(x)} \cdot X(\varphi(x)) =
b(\varphi(x)),
$$
so we have
\beasn
\addtocounter{equation}{-1}
\setcounter{subequation}{2}
&& \Phi_x \cdot b(x) = b(\varphi(x)) ,
\label{sym3'}
\eeasn
which means that $\Phi_*[b] = b$.

Conversely, from $\Phi_*[A] = A$ we have the first of equations
(\ref{sym2}), and, from $\Phi_*[b] = b$,
$$
A_{\varphi(x)} \cdot \left(\Tan_x(\varphi) \cdot X(x) - X(\varphi(x))\right) =
\Phi_x \cdot A_x \cdot X(x) - A_{\varphi(x)} \cdot X(\varphi(x)) =
\Phi_x \cdot b_x - b_{\varphi(x)} = 
0 ,
$$
therefore we have the second of equations (\ref{sym2}).
\qed

A final remark: as before, $\Phi$ is uniquely defined on 
$\Img A|_{M_1}$.

\section{Infinitesimal symmetries}

The infinitesimal version of an automorphism of a differential manifold
is a vector field~$X$,
in the sense that integration of it yields a local 1-parametre
group of diffeomorphisms, the flow $\Fl_X$.
We can translate the results of the preceding section to the
infinitesimal language.
The basic geometric tools are gathered in the appendix.

\begin{defin}
An {\it infinitesimal symmetry}\/ of an implicit system 
$D \subset \Tan M$ is a vector field $V$ on~$M$ such that 
the maps $\Fl_V^\eps$ are locally symmetries of~$D$. 
\end{defin}

\begin{prop}
$V$ is an infinitesimal symmetry of~$D$ iff 
its canonical lift to $\Tan M$, $V^\Tan$, is tangent to~$D$. 
\end{prop}
\proof 
It follows from the definition of infinitesimal symmetry 
and from the definition of the vector field $V^\Tan$, 
whose flow is constituted by the maps $\Tan(\Fl_V^\eps)$. 
\qed

\begin{defin}
An {\it infinitesimal symmetry}\/ of a linearly singular system 
$(A \colon \Tan M \to F,b)$
is an infinitesimal automorphism $(V,W)$ of the vector bundle 
$\pi \colon F \to M$
such that its flow $(\Fl_V^\eps,\Fl_W^\eps)$ is constituted by
local symmetries of the linearly singular system.
\end{defin}

According to proposition~\ref{prop-vectaut} in the appendix,
being $(V,W)$ 
an infinitesimal automorphism of vector bundles means that
$V$ is a vector field on~$M$,
$W$ is a vector field on~$F$, 
and $W$ projects to~$V$ and is a linear vector field.

\begin{teor}
An infinitesimal vector bundle automorphism $(V,W)$ 
is an infinitesimal symmetry of the system iff
\beq
\Tan b \comp V = W \comp b ,
\label{infsym-b}
\eeq
\beq
\Tan A \comp V^\Tan = W \comp A .
\label{infsym-A}
\eeq
\label{teor-infsym}
\end{teor}
\proof
The conditions (\ref{mor}) for the couple of flows $(\Fl_V^\eps,\Fl_W^\eps)$ 
to be a symmetry may be written as 
$$
b = \Fl_W^{-\eps} \comp b \comp \Fl_V^{\eps} ,
\qquad
A = \Fl_W^{-\eps} \comp A \comp \Tan \Fl_V^{\eps} .
$$
According to proposition~\ref{prop-invar} in the appendix, 
these equalities hold for each~$\eps$ iff 
(\ref{infsym-b}) and (\ref{infsym-A}) also do.
\qed

Now we can state the infinitesimal version of theorem~\ref{teor-main}:
 
\begin{teor}
Let $(A \colon \Tan M \to F,b)$ be a linearly singular system 
satisfying the regularity assumption,
and let $D = A^{-1}(b(M))$ be the associated implicit system.

A vector field $V$ on~$M$ is an infinitesimal symmetry of~$D$
iff, 
locally, there exists a vector field $W$ on~$F$ such that
$(V,W)$ is an infinitesimal symmetry of the linearly singular system.
\label{teor-infmain}
\end{teor}
\proof
Consider local coordinates $(x^i)$ on~$M$, 
$(x^i,u^i)$ on $\Tan M$ and $(x^i,v^k)$ on~$F$.
Then the section $b$ reads $(x^i) \mapsto (x^i,b^k(x))$,
and the morphism~$A$ reads $(x^i,u^i) \mapsto (x^i,A^k_{\,i}(x)u^i)$.
Let us write 
$$
V =  a^i \derpar{}{x^i} ,
$$
so that
$\ds
V^\Tan = a^i \derpar{}{x^i} + \derpar{a^i}{x^j}u^j \derpar{}{u^i}
$.
The subset $D \subset \Tan M$ is locally defined by the vanishing of
the constraints $\psi^k(x,u) := A^k_{\,i}(x) u^i - b^k(x)$.
In this way, the tangency of $V^\Tan$ to $D$ is locally expressed as
$$
V^\Tan \lop \psi^k = B^k_{\,l} \psi^l ,
$$
for some functions $B^k_{\,l}$;
in principle,
these functions depend on $(x^i,u^i)$,
and may not be unique.
However, 
the derivative of an affine function with respect to a linear vector field
is again an affine function,
so the functions $B^k_{\,l}$ can be assumed not to depend on~$u^i$.
Writing more explicitly the preceding equality we obtain
$$
A^k_{\,j} \derpar{a^j}{x^i} u^i +
\derpar{A^k_{\,i}}{x^j} a^j u^i -
\derpar{b^k}{x^i} a^i
=
B^k_{\,l} A^l_{\,i} u^i - B^k_{\,l} b^l .
$$
Equating the constant and the linear parts,
we conclude that $V$ is an infinitesimal symmetry of~$D$ iff 
there are functions $B^k_{\,l}(x)$ such that
\beq
\qquad
\derpar{f^k}{x^i} a^i
=
B^k_{\,l} f^l ,
\qquad
A^k_{\,j} \derpar{a^j}{x^i} +
\derpar{A^k_{\,i}}{x^j} a^j 
=
B^k_{\,l} A^l_{\,i} .
\label{Vsymm-c}
\eeq

On the other hand,
a linear vector field $W$ on $F$ projecting to~$V$ is expressed as
$$
W = a^i \derpar{}{x^i} + B^k_{\,l}(x)v^l \derpar{}{v^k} 
$$
for some other functions $B^k_{\,l}$.
Then, according to theorem~\ref{teor-infsym}, 
the conditions of being $(V,W)$ an infinitesimal symmetry of  
$(A \colon \Tan M \to F,b)$
read as follows: 
$\Tan b \comp V = W \comp b$ means
$$
\derpar{b^k}{x^i} a^i = B^k_{\,l} b^l ,
$$
and
$\Tan A \comp V^\Tan = W \comp A$ means
$$
\left( \derpar{A^k_{\,i}}{x^j} a^j + A^k_{\,j} \derpar{a^j}{x^i} \right)u^i =
B^k_{\,l} A^l_{\,i} .
$$
Comparing these conditions with (\ref{Vsymm-c}),
we conclude that being $V$ an infinitesimal symmetry of~$D$
is equivalent to the existence of $W$ making
$(V,W)$ an infinitesimal symmetry of $(A \colon \Tan M \to F,b)$.
\qed

\section{Regular systems and consistent systems}

\subsection{Regular systems}

\begin{defin}
A linearly singular system $(A \colon \Tan M \to F,b)$ 
is {\it regular}\/ if 
$A$ is a vector bundle isomorphism. 
\end{defin}

In this case the dynamics is uniquely determined
by the associated {\it explicit system}\/ 
given by the vector field $X=A^{-1} \comp b$.

Note that, for a diffeomorphism $\varphi \colon M \to M$,
now it is equivalent to say 
that $D = X(M)$ is invariant by $\Tan \varphi$, 
and that $X$ is invariant by~$\varphi$.

If $(\varphi,\Phi)$ is a symmetry of the linearly singular system,
then $\Phi$ is uniquely determined from $\varphi$ as
\beq
\Phi = A \comp \Tan \varphi \comp A^{-1} .
\label{Phi}
\eeq
Then the relations
$\Phi \comp b = b \comp \varphi$ and
$\Tan \varphi \comp X = X \comp \varphi$
are readily seen to be equivalent.
So we have proved the following result:

\begin{prop}
Suppose that the system $(A \colon \Tan M \to F,b)$ is regular, 
and let $\varphi \colon M \to M$ be a diffeomorphism.
Then the following statements are equivalent:
\ben
\itemsep 0pt plus 1pt
\item
$\varphi$ is a symmetry of the associated implicit system~$D$.
\item
$\varphi$ leaves the dynamical vector field $X$ invariant.
\item
$\varphi$ is the base map of a symmetry $(\varphi,\Phi)$ 
of the linearly singular system $(A \colon \Tan M \to F,b)$
---then $\Phi$ is uniquely determined by equation (\ref{Phi}).
\qed
\een
\end{prop}

As for the infinitesimal symmetries, we have a similar situation:
equation (\ref{infsym-A}) determines $W = A_*[V^\Tan]$,
and then equation (\ref{infsym-b}) says
$\Tan X \comp V = V^\Tan \comp X$,
which means $[V,X]=0$. 
So we have:

\begin{prop}
Suppose that the system $(A \colon \Tan M \to F,b)$ is regular, 
and let $V$ be a vector field in~$M$.
Then the following statements are equivalent:
\ben
\itemsep 0pt plus 1pt
\item
$V$ is an infinitesimal symmetry of 
the associated implicit system~$D$.
\item
$V$ leaves the dynamical vector field $X$ invariant ($[V,X]=0$).
\item
There exists an infinitesimal symmetry $(V,W)$ 
of the linearly singular system $(A \colon \Tan M \to F,b)$
---then $W$ is uniquely determined as $W = A_*[V^\Tan]$.
\qed
\een
\end{prop}

An important case of a regular system 
is that of a hamiltonian system $(M,\omega,H)$,
where $\omega$ is a symplectic form on a manifold~$M$ and 
$H$ is a hamiltonian function.
This defines a linearly singular system
$(\hat\omega \colon \Tan M \to \Tan^*M,\dif H)$,
whose dynamics is ruled by the hamiltonian vector field~$X_H$.
A diffeomorphism $\varphi$ preserving $X_H$ 
is not in general a symmetry of the hamiltonian system,
since it may not be a canonical transformation
(symplectomorphism).
However, we have shown that it defines a symmetry of the system 
if considered as a linearly singular system.

\subsection{Consistent systems}

\begin{defin}
A linearly singular system $(A \colon \Tan M \to F,b)$
is {\it consistent}\/ if
$A$ is a surjective vector bundle morphism.
\end{defin}

The solutions of the corresponding differential equation are 
the integral curves of the primary vector fields, that is, 
the sections of the affine bundle $D \to M$;
they can be expressed as 
$X_o + \Gamma$,
where $X_o$ is a particular primary vector field 
and $\Gamma$ belongs to $\Ker A$.
So, the invariance of~$D$ can be stated in terms of $X_o$ and $\Ker A$:

\begin{prop}
Let $(A \colon \Tan M \to F,b)$ be a 
consistent linearly singular system,
and let $D$ be the associated implicit system.
\ben
\item
A diffeomorphism $\varphi \colon M \to M$ is a symmetry of~$D$ iff
$\Ker A$ is invariant by $\varphi$ and,
for any primary vector field $X_o$,
$\varphi_*[X_o]-X_o$ is in $\Ker A$.
\item
A vector field $V$ in~$M$ is an infinitesimal symmetry of~$D$ iff,
for any vector field $\Gamma$ in $\Ker A$,
$[V,\Gamma]$ is in $\Ker A$, and,
for any primary vector field $X_o$,
$[V,X_o]$ is in $\Ker A$. 
\een 
\end{prop} 
\proof 
The first statement follows from the equivalence between assertions 1 and~2 
in theorem~\ref{teor-main}. 
The second statement is the infinitesimal version of the first one.
\qed 

Note that a consistent system is locally solvable, 
so the preceding conditions characterise the
dynamic symmetries of the linearly singular differential equation.

Let us remark that the definition of a consistent system 
could be slightly more general. 
If the consistency condition for the linear equation
$A_x \lop u_x = b(x)$ holds at every $x \in M$
({\it i.e.}, $M_1 = M$),
then the image of $b$ is contained in $\Img A \subset F$.
If this is a subbundle, we could safely substitute $\Img A$ for $F$
in the linearly singular system,
thus obtaining what we have called a consistent system.
A similar remark can be applied to regular system.

\subsection{Symmetries at the end of the constraint algorithm}

Consider a linearly singular system 
$(A \colon \Tan M \to F,b)$.
If the constraint algorithm as explained in section~2 
can be performed on it 
(in the sense that 
the regularity assumption is satisfied at each step of the algorithm), 
then the final dynamics is that of a 
consistent linearly singular system, 
$(A_\fin \colon \Tan M_\fin \to \Img A_\fin,b_\fin)$,
so the preceding proposition may be directly applied to it.

\section{An example: the associated presymplectic system}

In some problems of control theory, 
equations of the type $A(x)\dot x = b(x,u)$,
where $u$ represents the control, 
play a relevant role. 
Ibort noted 
---see
\cite{DI}---
that, 
from a linearly singular system 
$(A \colon \Tan M \to F,b)$, 
one can define a presymplectic system $(F^*,\omega,H)$  
on the total space of the dual vector bundle 
$\pi^* \colon F^* \to M$. 
This is as follows.

If $\theta_M$ is the canonical 1-form 
and $\omega_M = -\dif \theta_M$ is the canonical symplectic form
of $\Tan^*M$, 
one can use the transpose map $\transp{A} \colon F^* \to \Tan^*M$ of~$A$ 
to define forms on $F^*$
by pull-back:
\beq
\theta = \transp{A}^*[\theta_M] ,
\qquad
\omega = \transp{A}^*[\omega_M] .
\eeq
In a similar way one can use the section~$b \colon M \to F$ to define 
a linear function $H \colon F^* \to \Real$:
\beq
H(\alpha_x) = \langle \alpha_x , b(x) \rangle .
\eeq

One can study the relations between both systems.
For instance,
each solution of the linearly singular equation 
is in correspondence with
a family of solutions of the equation of motion of the presymplectic system.
We shall limit ourselves to study the relation between 
the symmetries of both systems.

Consider a vector bundle automorphism 
$(\varphi,\Phi)$ of $\pi \colon F \to M$.
The contragradient map $\Phi^\vee = \transp{\Phi}^{-1}$
is a vector bundle automorphism of $F^*$,
with base map~$\varphi$.

In analogy to (\ref{push}), $\Phi^\vee$ transforms 
the 1-form $\theta$ and the hamiltonian function:
$$
\Phi^\vee_*[\theta] =
 (\Tan \Phi^\vee)^\vee \comp \theta \comp (\Phi^\vee)^{-1} ,
\qquad
\Phi^\vee_*[H] = H \comp (\Phi^\vee)^{-1} .
$$

Note that $H = \langle \Id , b \comp \pi^* \rangle$. 
A computation shows that
\beq
\Phi^\vee_*[H] = \langle \Id , \Phi_*[b] \comp \pi^* \rangle
\eeq
which proves that if $b$ is $\Phi$-invariant
then $H$ is $\Phi^\vee$-invariant;
the converse is also true, since $b$ is determined by~$H$.

In a similar way, consider for the sake of simplicity 
the 1-form $\theta$ as a linear function 
$\theta \colon \Tan F^* \to \Real$ 
---we use the same letter $\theta$ to not overload the notations.
Then note that $\theta = \langle \tau_{F^*} , A \comp \Tan \pi^* \rangle$.
Another computation shows that
\beq
\Phi^\vee_*[\theta] = 
\langle \tau_{F^*} , \Phi^\vee_*[A] \comp \Tan \pi^* \rangle .
\eeq
As before, this proves that 
$A$ is $\Phi$-invariant iff $\theta$ is $\Phi^\vee$-invariant.

As a conclusion, we have:
\begin{prop}
Consider a linearly singular system 
$(A \colon \Tan M \to F,b)$, 
and the associated presymplectic system
$(F^*,-\dif \theta,H)$.
Let $(\varphi,\Phi)$ be a vector bundle automorphism of $F \to M$.
Then:
\ben
\itemsep 0pt plus 1pt
\item
$b$ is $\Phi$-invariant iff $H$ is $\Phi^\vee$-invariant.
\item
$A$ is $\Phi$-invariant iff $\theta$ is $\Phi^\vee$-invariant.
\een
So if $(\varphi,\Phi)$ is a symmetry of the linearly singular system 
then 
$\Phi^\vee$ is a symmetry of the presymplectic system.
\qed
\end{prop}

\section{Conclusions}

In this paper we have studied the symmetries of
linearly singular differential equations. 
To do this,  
we have intended to clarify what a ``symmetry'' is for a differential
equation, 
and we have found that there are several legitimate approaches to this concept. 
Some approaches rely on the geometry of the data defining 
the differential equation (``geometric symmetries''), 
whereas others simply characterise the symmetry as a transformation that 
maps solutions into solutions (``dynamic symmetries''). 
As for the type of symmetries, 
we have considered only point symmetries, 
which are those defined by diffeomorphisms of the configuration space 
(or vector fields, in the infinitesimal case).

After a general introduction to
linearly singular differential equations,
we deal with two different concepts of geometric symmetry. 
The first one, the symmetry of an implicit system, 
has a general applicability and relies on the invariance of 
the subset $D \subset \Tan M$ that defines, implicitly, 
the differential equation.
The second one, the symmetry of a linearly singular system, 
is specific of the systems discussed in this paper. 
For them, 
we show that both concepts of symmetry are essentially equivalent. 
We prove it for a general, finite, transformation, and also for the case
of flows generated by infinitesimal transformations.

Under appropriate regularity assumptions, 
the concept of dynamic symmetry of the implicit differential equation 
defined by $D \subset \Tan M$ 
is equivalent 
to the concept of geometric symmetry 
of an appropriate locally solvable system $D' \subset \Tan M'$; 
this also holds for the linearly singular case, 
by means of the constraint algorithm. 
We show that, under appropriate regularity assumptions,  
the most general dynamic symmetry for a linearly singular equation 
(on a manifold~$M$)
can be locally realised as a geometric symmetry of a linearly singular system 
(on the final constraint manifold~$M_\fin$).

As for the tools needed to deal with the infinitesimal symmetries,
in the appendix
we have performed a careful study of infinitesimal transformations
and the invariance of geometric structures under the action of flows.
This has been applied to infinitesimal automorphisms of vector bundles, 
and may be useful to deal with other problems about infinitesimal invariance.

\appendix
\section*{Appendix: calculus with infinitesimal variations} 
\stepcounter{section}

\subsection*{General aspects}

To deal with infinitesimal symmetries it will be convenient
to perform a more general study.
Let us consider a map $f \colon \Real \times M \to N$.
We will use the notation 
$f_\eps(x) = f(\eps,x)$,
and we can interpret the maps $f_\eps \colon M \to N$ as 
a ``variation'' of the map~$f_o = f(0,-)$.
The corresponding ``infinitesimal variation'' of~$f$ is 
the map $\w_f \colon M \to \Tan N$ defined as
\beq
\w_f (x) = \restric{\derpar{}{\eps}}{\eps=0} f ;
\label{vari}
\eeq
in other words, it is $\w_f(x) = f'(0,x)$,
where
$$
f' = \Tan f \comp E \colon \,\Real \times M \to \Tan N
$$
is the $\eps$-derivative of~$f$
---here $E$ is the unit vector field on~$\Real$ interpreted as 
to be on $\Real \times M$.


Note that $\w_f$ is a vector field along $f_o$.
(Interpreting $f$ as a path in the infinite-dimensional manifold
of maps from $M$ to~$N$, $\w_f$ is its tangent vector at $\eps=0$.)
It is clear that if $f_\eps$ does not depend on~$\eps$,
then $\w_f = 0$
(the converse is obviously not true).

Given a number $c$, if we define $g(\eps,x) = f(c\eps,x)$
it is easily seen that $\w_g = c \w_f$,
and in particular the infinitesimal variation of $f(-\eps,x)$ is 
$-\w_f$.

Now let us consider another variation
$g \colon \Real \times N \to P$,
and construct the compositions $g_\eps \comp f_\eps$;
they define a variation
$g \scomp f \colon \Real \times M \to P$:
\beq
(g \scomp f)(\eps,x) = g(\eps,f(\eps,x)) .
\label{comp}
\eeq
A direct computation in coordinates shows that
\beq
\w_{g \scomp f} = \w_g \comp f_o + \Tan g_o \comp \w_f .
\label{varicomp}
\eeq
Let us show these objects in a diagram:
$$
\diagr(2,1.15) 
\capsa(0,0){M} 
\capsa(0,0){\flE{f_o}} 
\capsa(0,0){\flNE{\w_f}}
\capsa(1,0){N} 
\capsa(1,1){\Tan N} 
\capsa(1,1){\flS{}}
\capsa(1,0){\flE{g_o}} 
\capsa(1,0){\flNE{\w_g}}
\capsa(1,1){\flE{\Tan g_o}}
\capsa(2,0){P} 
\capsa(2,1){\Tan P} 
\capsa(2,1){\flS{}}
\enddiagr 
$$
\noindent
This result is immediately extended to the composition of
three (or more) variations.

Another immediate consequence is the following one.
Suppose that the $f_\eps \colon M \to N$ are diffeomorphisms, 
and set $g_\eps = f_\eps^{-1}$.
Then we have
$$
\w_g = - \Tan g_o \comp \w_f \comp g_o .
$$

Finally, in a similar way one can prove that, 
if $h_\eps = (f_\eps,f'_\eps)$
then $\w_h = (\w_f,\w_{f'})$,
with the usual identification $\Tan(N \times N') = \Tan N \times \Tan N'$.

\subsection*{Transformation of maps}

Consider variations 
$f_\eps \colon M \to M'$ and
$g_\eps \colon N \to N'$,
and a map $h' \colon M' \to N'$.
If the $g_\eps$ are diffeomorphisms,
we can construct a family of maps $h_\eps \colon M \to N$ as
$$
h_\eps = g_\eps^{-1} \comp h' \comp f_\eps ,
$$
for which one can easily compute~$\w_h$.

In the important particular case where $M=M'$, $N=N'$, and
$f$ and~$g$ are variations of the identity, 
obviously one has that $h_o = h'$ and 
\beq
\w_h = -\w_g \comp h_o + \Tan h_o \comp \w_f .
\label{varifun}
\eeq

Now let us take fibre bundles 
$\pi \colon M \to B$ and $\rho \colon N \to C$.
Suppose that the couple
$g_\eps \colon M \to N$ and $f_\eps \colon B \to C$
is a variation of a bundle morphism
$(g_o,f_o)$.
If $(g_\eps,f_\eps)$ are bundle morphisms 
({\it i.e.}, $\rho \comp g_\eps = f_\eps \comp \pi$)
then
\beq
\Tan \rho \comp \w_g = \w_f \comp \pi .
\eeq
One can consider more particularly the case where
$g$ and~$f$ are variations of the identity,
and ask for instance whether a section 
$\sigma \colon B \to M$ 
is transformed into sections, 
or is left invariant, etc.

\subsection*{Variations defined in terms of flows}

Note that, with slight complications,
the preceding definitions and results could be applied
to variations $f$ defined only on an open subset 
$D \subset \Real \times M$ containing $\{0\} \times M$.
This remark is especially relevant for what follows.

Let us consider a vector field $X$ on~$M$,
and denote its flow by $\Fl_X$,
so that the integral curve with initial condition~$x$
is $\eps \mapsto \Fl_X^\eps(x)$.
Let us assume for simplicity that $X$ is complete, 
that is, the domain of $\Fl_X$ is $\Real \times M$; 
otherwise we would apply the preceding remark.
It is clear from the definition of the flow that
\beq
\Fl_X' = X \comp \Fl_X ,
\label{derflow}
\eeq
$$
\Fl_X^0 = \Id, 
\qquad
\w_{\Fl_X} = X .
$$
Note that applying (\ref{derflow})
one can compute the $\eps$-derivative of the expression
$F^\eps(F^{-\eps}(x))=x$ 
and obtain 
\beq
\Tan_{x} \Fl_X^{\eps} \lop X(x) = X(\Fl_X^\eps(x)) ,
\label{flowinv}
\eeq
which indeed tells that $X$ is invariant under its flow.

Now let us consider a map $h_o \colon M \to N$,
and vector fields $X$ on~$M$ and $Y$ on~$N$.
We can use their flows to transform $h_o$ as
\beq
h_\eps(x) = h(\eps,x) = \Fl_Y^{-\eps} (h_o(\Fl_X^{\eps}(x))) .
\label{flowtrans}
\eeq
Applying (\ref{varifun}) to this composition we obtain
\beq
\w_h = \Tan h_o \comp X - Y \comp h_o .
\label{Lie-h}
\eeq
Note therefore that $\w_h$ is zero
iff the vector fields $X$ and $Y$
are $h_o$-related.

In 
\cite{KMS-natural}
such infinitesimal variations are studied under the name of
``generalised Lie derivatives'',
and for instance $\w_h$ in (\ref{Lie-h}) is denoted by
$\tilde\Lie_{(X,Y)} h_o$.

Let us show all these objects in a diagram:
$$
\diagr(3,1.15) 
\capsa(0,0){M} 
\capsa(0,1){M} 
\capsa(1,0){N} 
\capsa(1,1){N} 
\capsa(0,1){\flE{h_\eps}} 
\capsa(0,1){\flS*{\Fl_X^\eps}}
\capsa(0,0){\flE{h_o}} 
\capsa(1,0){\flN*{\Fl_Y^{-\eps}}} 

\capsa(2,0){M} 
\capsa(2,1){\Tan M} 
\capsa(2,0){\flN{X}}
\capsa(2,0){\flE{h_o}} 
\capsa(2,0){\flNE{\w_h}}
\capsa(2,1){\flE{\Tan h_o}}
\capsa(3,0){N} 
\capsa(3,1){\Tan N} 
\capsa(3,0){\flN{Y}}
\enddiagr 
$$

\begin{prop}
With the previous notations,
$h_o$ is invariant under the action of the couple of flows 
iff $\w_h = 0$.
\label{prop-invar}
\end{prop}

\proof
The direct implication is a trivial consequence of the definition of~$\w_h$.
For the converse, 
to show that $h(\eps,x) = h(0,x)$
we will prove that the $\eps$-derivative of $h$, $h'(\eps,x)$, 
vanishes at any~$\eps$
(not only at $\eps=0$).
This will be a consequence of the following equation:
\beq
h'(\eps,x) = 
\Tan_{h_o(\Fl_X^\eps(x))} \Fl_Y^{-\eps} \lop 
(\Tan h_o \comp X - Y \comp h_o)(\Fl_X^\eps(x)) .
\label{eqfon}
\eeq
To prove it, let us introduce some notation.
First, we write $F$ and $G$ the flows of $X$ and~$Y$.

The tangent map of~$F$ applied to a couple of vectors
$((\eps,\tau),u_x) \in \Tan_\eps \Real \times \Tan_x M \isom
\Tan_{(\eps,x)}(\Real \times M)$ can be written
\beq
\Tan F ((\eps,\tau),u_x) = F'(\eps,x)\tau + \Tan_x F^\eps \lop u_x .
\label{tanfflow}
\eeq
Changing some terms in (\ref{flowinv}), 
we also have 
\beq
Y(y) = \Tan_{G^\eps(y)} G^{-\eps} \lop Y(G^\eps(y)) .
\label{Yflow}
\eeq

Now let us proceed to compute $h' = \Tan h \comp E$.
Let us write $h = \bar G \comp \bar h_o \comp \bar F$,
where 
$\bar F(\eps,x) = (\eps,F(\eps,x))$,
$\bar h_o(\eps,x) = (\eps,h_o(x))$, and
$\bar G(\eps,x) = G(-\eps,x)$.
The chain rule applied to these maps,
as well as (\ref{tanfflow}) and (\ref{derflow}), yield
$$
h'(\eps,x) = 
-Y(G^{-\eps}(h_o(F^\eps(x)))) + 
\Tan_{h_o(F^\eps(x))} G^{-\eps} \lop \Tan_{F^\eps(x)} h_o \lop X(F^\eps(x)) .
$$
Application of (\ref{Yflow}) converts this equation into
\beann
h'(\eps,x) &=& 
-\Tan_{h_o(F^\eps(x))} G^{-\eps} \lop Y(h_o(F^\eps(x))) + 
\Tan_{h_o(F^\eps(x))} G^{-\eps} \lop \Tan_{F^\eps(x)} h_o \lop X(F^\eps(x)) .
\\
&=& 
\Tan_{h_o(F^\eps(x))} G^{-\eps} \lop
\left(
  \Tan_{F^\eps(x)} h_o \lop X(F^\eps(x)) - (Y \comp h_o)(F^\eps(x)) 
\right) 
\\
&=& 
\Tan G^{-\eps} \comp
\left( \Tan h_o \comp X - Y \comp h_o \right)
(F^\eps(x)) ,
\\
&=& 
(\Tan G^{-\eps} \comp \w_h)
(F^\eps(x)) ,
\eeann
which is (\ref{eqfon}).
\qed

\subsection*{Infinitesimal vector bundle automorphisms}

Let $\pi \colon E \to M$ be a fibre bundle.
An {\it infinitesimal automorphism}\/
is a couple of vector fields $(X,Y)$ on $M$ and $E$
such that their flows $(F^\eps,G^\eps)$
are fibre bundle isomorphisms.

\begin{prop}
With the preceding notations,
$(X,Y)$ is an infinitesimal automorphism iff
$Y$ is projectable to~$X$.
\label{prop-fibaut}
\end{prop}

\proof
The condition of being morphism is 
$\pi \comp G^\eps = F^\eps \comp \pi$,
that is to say,
$\pi = F^{-\eps} \comp \pi \comp G^\eps$.
According to proposition~\ref{prop-invar},
the right-hand side is $\eps$-invariant iff
$\Tan \pi \comp Y = X \comp \pi$,
which is the condition of projectability.
\qed

Note that under these conditions 
the vector field $X$ is determined by~$Y$,
so we can as well say that $Y$ is an infinitesimal automorphism
of the fibre bundle.

From now on let us suppose that $\pi \colon E \to M$ is a {\it vector}\/ bundle.
A couple of vector fields $(X,Y)$ 
is an {\it infinitesimal vector bundle automorphism}\/
if their flows $(F^\eps,G^\eps)$
are vector bundle isomorphisms.
Of course $Y$ is projectable to~$X$.
$Y$ is said to be a {\it linear vector field} if
$(X,Y)$ is a morphism between the vector bundles 
$\pi \colon E \to M$ and $\Tan \pi \colon \Tan E \to \Tan M$:
$$
\diagr[10ex](1,1.2) 
\capsa(0,0){M} 
\capsa(0,1){E} 
\capsa(1,0){\Tan M} 
\capsa(1,1){\Tan E} 
\capsa(0,1){\flE{Y}} 
\capsa(0,1){\flS*{\pi}}
\capsa(0,0){\flE{X}} 
\capsa(1,1){\flS{\Tan \pi}} 
\enddiagr 
$$

\begin{prop}
Let $Y$ be a $\pi$-projectable vector field on~$E$.
Then $Y$ is an infinitesimal vector bundle automorphism
iff $Y$ is a linear vector field.
\label{prop-vectaut}
\end{prop}

\proof
First remember that a smooth function defined on a vector space
is homogeneous of degree one iff it is linear.

So, denoting by $m_\lambda \colon E \to E$
the multiplication by~$\lambda$,
the fact that $G^\eps$ is linear is equivalent to
$G^\eps \comp m_\lambda = m_\lambda \comp G^\eps$
for each~$\lambda$.
This can be expressed also as
$m_\lambda = G^{-\eps} \comp m_\lambda \comp G^\eps$,
and applying proposition~\ref{prop-invar} again, 
this holds for each~$\eps$ iff
$\Tan m_\lambda \comp Y = Y \comp m_\lambda$.
Written in other terms,
we have
$\lambda \lop Y(u) = Y(\lambda u)$,
where the first product is meant to be with respect to 
the vector bundle structure
$\Tan \pi \colon \Tan E \to \Tan M$.
So on each fibre $Y_x \colon E_x \to \Tan_{X(x)}E$ is 
a homogeneous smooth map,
and therefore linear.
\qed

A different proof of this result, using coordinates, can be found in
\cite{KMS-natural}.
It is also interesting to recall that
a projectable vector field $Y$ is linear iff
it is invariant under the action of the Liouville's vector field,
$[\Lio_E,Y] = 0$.

\subsection*{Acknowledgements}

X.\,G. acknowledges financial support by CICYT project 
PB98--0920.
J.\,M.\,P. acknowledges financial support by 
CICYT, AEN98-0431, and CIRIT, GC 1998SGR.



\end{document}